\documentclass[aps,prd,reprint,longbibliography,titlepage,nofootinbib]{revtex4-1}

\usepackage{amsthm}
\usepackage{amssymb}   
\usepackage{dsfont}
\usepackage{mathtools} 
\usepackage{hyperref}
\hypersetup{colorlinks=true,linkcolor=blue,urlcolor=blue,citecolor=blue}
\usepackage{accents}
\usepackage{tensor}
\usepackage{xcolor}
\usepackage{graphicx}
\usepackage[cal=boondox]{mathalfa}
\usepackage{lipsum}
\usepackage{relsize}
\usepackage{comment}
\usepackage{nameref}
\newcommand{\uac}[1]{\underaccent{\tilde}{#1}}
\newcommand{\uacc}[1]{\uac{\uac{#1}}}

\begin{document}
	
	\title{Linking the ADM formulation to other Hamiltonian formulations of general relativity}

	\author{Merced Montesinos$\,$\href{https://orcid.org/0000-0002-4936-9170} {\includegraphics[scale=0.05]{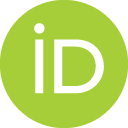}}}
	\email[Corresponding author \\]{merced@fis.cinvestav.mx}
	\affiliation{Departamento de F\'{i}sica, Centro de Investigaci\'on y de Estudios Avanzados del Instituto Politécnico Nacional, Avenida Instituto Polit\'{e}cnico Nacional 2508,\\
		San Pedro Zacatenco, 07360 Gustavo Adolfo Madero, Ciudad de M\'{e}xico, Mexico}
	
	\author{Jorge Romero$\,$\href{https://orcid.org/0000-0001-8258-6647} {\includegraphics[scale=0.05]{ORCIDiD_icon128x128.png}}}
	\email{jorge.romero@correo.nucleares.unam.mx}
	\affiliation{Departamento de F\'{i}sica de Altas Energ\'{i}as, Instituto de Ciencias Nucleares, Universidad Nacional Aut\'{o}noma de M\'{e}xico, Apartado Postal 70-543, Ciudad de M\'{e}xico, 04510, Mexico
}
	
	\date{\today}
	
	\begin{abstract}
	We obtain the Arnowitt-Deser-Misner formulation of general relativity in $n$ dimensions ($n \geq 3$) from its either $SO(n-1,1)$ [$SO(n)$] or  $SO(n-1)$ Palatini Hamiltonian formulations and vice versa [we recall that $SO(n-1,1)$ [$SO(n)$] requires no gauge fixing whereas $SO(n-1)$ involves the time gauge]. Similarly, the Hamiltonian formulation of general relativity in terms of Ashtekar-Barbero variables can also be directly obtained from the Arnowitt-Deser-Misner Hamiltonian formulation and vice versa, which is an alternative approach to the way followed by Barbero. We give the relevant maps among the phase-space variables and relate the corresponding symplectic structures and the first-class constraints.
	\end{abstract}
	
	\maketitle

	
	\section{Introduction}\label{intro}
	The Hamiltonian analysis of general relativity has been a fruitful arena of research since the pioneering work of Arnowitt, Deser, and Misner (ADM)~\cite{ADM}. The usual approach to get the Hamiltonian formulation of general relativity---and, in fact, of any other theory---is to begin with a particular Lagrangian formulation of it~\cite{DiracBook}. This was indeed the route followed by ADM. Another way is to perform a canonical transformation from a given Hamiltonian formulation, which was the approach followed by Barbero~\cite{Barbero9505} to get the Hamiltonian formulation of general relativity that is the starting point of the quantum theory known as loop quantum gravity~\cite{RovBook,ThieBook}.
	
	Nevertheless, there are other approaches to get a Hamiltonian formulation of general relativity that do not follow the usual or standard Dirac approach, among them we find the seminal work of Ashtekar~\cite{Ashtekar8611} and also the work of Thiemann~\cite{ThieBook}, which link, respectively, the ADM formulation with the Ashtekar and $SO(n-1)$ formulations of general relativity. The relevance of these two latter works is that they both link Hamiltonian formulations of general relativity that have a different number of phase-space variables. In this paper, we follow this line of thought and relate the ADM formulation in $n$ dimensions ($n \geq 3$) with the $SO(n-1,1)$ [$SO(n)$] Hamiltonian formulation of general relativity that keeps intact the local $SO(n-1,1)$ [$SO(n)$] invariance and involves only first-class constraints~\cite{Bodendorfer1301b,Montesinos2001}. The link is made through two approaches: the top-down, in which we begin with the $SO(n-1,1)$ [$SO(n)$] Hamiltonian formulation and get the ADM formulation (Sec.~\ref{Lorentz_to_ADM}); and the bottom-up, in which we begin with the ADM formulation and get the $SO(n-1,1)$ [$SO(n)$] Hamiltonian formulation (Sec.~\ref{ADM_to_Palatini}). As is clear after reading these two sections, the two approaches can be slightly modified to relate the ADM formulation in $n$ dimensions ($n \geq 3$) with the $SO(n-1)$ Hamiltonian formulation that is obtained when the time gauge is imposed in the $SO(n-1,1)$ [$SO(n)$] Hamiltonian formulation, which is reported in Sec.~\ref{vectors_ADM}. The spacetime of dimension four is particularly important. As we mentioned, one of the most relevant Hamiltonian formulations of general relativity in four dimensions is the one given in terms of Ashtekar-Barbero variables~\cite{Barbero9505} (see also Refs.~\cite{Holst9605,Barros0100,Montesinos1801,Montesinos1903,Montesinos2004a} for alternative ways of obtaining the Barbero formulation). Therefore, we report in Sec.~\ref{Ashtekar_Barbero} how to obtain directly the Barbero formulation from the ADM formulation (bottom-up approach) and, for the sake of completeness, how to obtain also the ADM formulation from the Barbero one (top-down approach). Finally, our concluding remarks are contained in Sec.~\ref{Sec_concl}. 
	
	Our conventions and notation are those of Refs.~\cite{Montesinos2001,Montesinos2004a}. In particular, we assume that the $n$-dimensional spacetime $M$ can be foliated by spacelike leaves diffeomorphic to $\Sigma$ so that $M$ is diffeomorphic to $\mathbb{R} \times \Sigma$, with $\Sigma$ being an orientable $(n-1)$-dimensional spatial manifold without boundary. We use local coordinates adapted to the foliation of spacetime, $M = \mathbb{R} \times \Sigma$, and so $t$ and $x^a$ ($a,b,c, \dots$ taking on the values $1, \ldots, n-1$) label the points on $\mathbb{R}$ and $\Sigma$, respectively. The indices $I,J,K, \ldots$ that take on the values $0,1, \ldots, n-1$ are $SO(n-1,1)$ [$SO(n)$] valued and are lowered and raised with the $n$-dimensional Minkowski ($\sigma=-1)$ or Euclidean ($\sigma=1$) metric $(\eta_{IJ}) = \mbox{diag} (\sigma, 1, \ldots, 1)$. Symmetrization or antisymetrization symbol in any couple of indices involves a $1/2$ factor. 
	
	\section{From $SO(n-1,1)$ [$SO(n)$] phase-space variables to ADM variables}\label{Lorentz_to_ADM}
	
	We begin with the Hamiltonian formulation of the $n$-dimensional Palatini action reported in Refs.~\cite{Bodendorfer1301b,Montesinos2001}, which involves the canonical pair $(Q_{aI}, \tilde{\Pi}^{aI})$ and is given by 
		\begin{eqnarray}
		\label{S_QP}
		S &= & \int_{\mathbb{R} \times \Sigma} dt d^{n-1}x \Big( 2\tilde{\Pi}^{a I}\dot{Q}_{a I} -\lambda_{IJ} \tilde{\mathcal{G}}^{I J} \notag \\
		&& - 2 N^{a}\tilde{\mathcal{D}}_{a} - \underaccent{\tilde}{N} \tilde{\tilde{\mathcal{H}}} \Big),
	\end{eqnarray}
	where $\Sigma$ is a $(n-1)$-dimensional spacelike hypersurface without boundary and 
	\begin{subequations}
		\begin{eqnarray}
			\tilde{\mathcal{G}}^{IJ} &:=& 2 \tilde{\Pi}^{a[I} Q_{a}{}^{J]}, \label{Gauss1}\\
			\tilde{\mathcal{D}}_{a} &:=& 2\tilde{\Pi}^{bI} \partial_{[a} Q_{b]I} - Q_{aI} \partial_{b}\tilde{\Pi}^{bI}, \\
			\tilde{\tilde{\mathcal{H}}} &:=& - \sigma \tilde{\Pi}^{a I} \tilde{\Pi}^{b J}R_{abIJ} +  2 \tilde{\Pi}^{a[I}\tilde{\Pi}^{|b|J]}Q_{a I}Q_{b J} \notag \\
			&& + 2 \sigma  h^{1/(n-2)} \Lambda,
		\end{eqnarray}
	\end{subequations}
	are the Gauss, diffeomorphism, and Hamiltonian first-class constraints, respectively. Here, $R_{ab}{}^I{}_J = \partial_{a}\Gamma_{b}{}^{I}{}_{J} - \partial_{b} \Gamma_{a}{}^{I}{}_{J} + \Gamma_{a}{}^{I}{}_{K} \Gamma_{b}{}^{K}{}_{J} - \Gamma_{b}{}^{I}{}_{K} \Gamma_{a}{}^{K}{}_{J}$ is the curvature of the connection $\Gamma_a{}^I{}_J$ defined by $\nabla_{a}\tilde{\Pi}^{b I} :=  \partial_{a}\tilde{\Pi}^{b I} + \Gamma^{b}{}_{a c}\tilde{\Pi}^{c I} - \Gamma^{c}{}_{a c}\tilde{\Pi}^{b I} + \Gamma_{a}{}^{I}{}_{J}\tilde{\Pi}^{b J} = 0$ 
	with $\Gamma^c{}_{ab}= \Gamma^c{}_{ba}$ and $\Gamma_{a}{}^{I}{}_{J}=-\Gamma_{aJ}{}^{I}$. Furthermore, $h=\det(\tilde{\tilde{h}}^{ab})$ with $\tilde{\tilde{h}}^{ab}:= \eta_{IJ} \tilde{\Pi}^{aI} \tilde{\Pi}^{bJ}$, and so $h$ is a tensor density of weight $2(n-2)$, and $\Lambda$ is the cosmological constant. As usual, the dot ``$\cdot$'' stands for the partial derivative with respect to the coordinate time $t$, $\partial_t$, whereas $\partial_a = {\partial}/{\partial x^a}$.
	
	The canonical formulation~\eqref{S_QP} was obtained in Ref.~\cite{Montesinos2001} from the $n$-dimensional Palatini action by using a suitable parametrization of the vielbein $e^I$ and the $SO(n-1,1)$ [$SO(n)$] connection $\omega^I{}_J$. Alternatively, the formulation~\eqref{S_QP} can also be obtained by explicitly solving the irreducible second-class constraints of the theory that are equivalent to the original reducible second-class constraints of the $n$-dimensional Palatini theory~\cite{Bodendorfer1301a,Bodendorfer1301b}. 
	
	To perform the symplectic reduction means to go from the Hamiltonian description in terms of the phase-space variables $(Q_{aI}, \tilde{\Pi}^{aI})$ to the Hamiltonian formulation in terms of the ADM variables $(q_{ab}, {\tilde \pi}^{ab})$ by fixing the gauge freedom generated by the Gauss constraint $\tilde{\mathcal{G}}^{IJ}$. According to Dirac's theory of constrained systems, a gauge condition must be imposed by hand. All of this is what we call the top-down approach. 
	
	\subsection{The geometrical variables involved in the symplectic reduction}\label{reduction1}
	We define the variables $(q_{ab}, {\tilde \pi}^{ab})$ on $\Sigma$ by first defining the inverse metric $q^{ab}$ on $\Sigma$
	\begin{eqnarray}\label{definvq}
		q^{ab} &:=& h^{-1/(n-2)} \tilde{\tilde{h}}^{ab}.
	\end{eqnarray}
	Thus, the inverse of $q^{ab}$ is given by 
	\begin{eqnarray}\label{defADMq}
		q_{ab}= h^{1/(n-2)} \uacc{h}_{ab},
	\end{eqnarray}
	where $\uacc{h}_{ab}$ is the inverse of $\tilde{\tilde{h}}^{ab}$ and then $q_{ac} q^{cb}=\delta^b_a$. Therefore, $q_{ab}$ is defined only in terms of the momentum variable $\tilde{\Pi}^{aI}$.
	
	We also define
	\begin{eqnarray}\label{defADMp}
		\tilde{\pi}^{ab} &:=&  \tilde{k}^{ab} - q_{cd} \tilde{k}^{cd} q^{ab}
	\end{eqnarray}
	with 
	\begin{eqnarray}
		\label{k_def}
		\tilde{k}^{ab} &:=&  q^{c(a}\tilde{\Pi}^{b)I} Q_{cI}.
	\end{eqnarray}
	Thus, because of~\eqref{definvq} and~\eqref{defADMq}, both $\tilde{k}^{ab}$ and $\tilde{\pi}^{ab}$ are functions of the canonical variables $(Q_{aI}, \tilde{\Pi}^{aI})$. The explicit form of $\tilde{\pi}^{ab}$ in terms of the phase-space variables is given by
	\begin{eqnarray}\label{defADMpfinal}
		\tilde{\pi}^{ab} &=& h^{-1/(n-2)} \left( \tilde{\tilde{h}}^{c(a}\tilde{\Pi}^{b)I} Q_{cI} - \tilde{\tilde{h}}^{ab} \tilde{\Pi}^{cI} Q_{cI}\right).
	\end{eqnarray}
	Equations~\eqref{defADMq} and~\eqref{defADMpfinal} define a projection map $(Q_{aI}, \tilde{\Pi}^{aI}) \longmapsto (q_{ab}, {\tilde \pi}^{ab})$. It will be shown some lines below that $(q_{ab}, {\tilde \pi}^{ab})$ are indeed the ADM phase-space variables of general relativity. Equivalently, we can say that $(q_{ab}, \tilde{\pi}^{ab})$ label the gauge orbits on the constrained surface generated by the Gauss constraint of the Hamiltonian formulation of the Palatini action in $n$ dimensions given by~\eqref{S_QP}. 	
	
	Due to the fact that~\eqref{defADMpfinal} can be thought as $n(n-1)/2$ equations for $n(n-1)$ unknowns $Q_{aI}$, we can solve for $Q_{aI}$ and obtain
\begin{eqnarray}
\label{Q2pi}
Q_{aI} &=&  h^{1/(n-2)} H_{abcd} \tilde{\Pi}^{b}{}_{I} \tilde{\pi}^{cd}  \notag \\ 
&& +\dfrac{1}{2} \uacc{h}_{ab} \tilde{\Pi}^{b}{}_{J} \left(\eta_{IK}+ \sigma n_{I} n_{K} \right) \tilde{\Pi}^{JK}, 
\end{eqnarray}
where $\tilde{\Pi}^{IJ} =-\tilde{\Pi}^{JI}$ are $n(n-1)/2$ independent variables of weight $1$, $n_{I}$ is the internal vector defined in (9) of Ref.~\cite{Montesinos2001} ($n_I n^I =\sigma$ and $n_I \tilde{\Pi}^{aI}=0$), and 
			\begin{eqnarray}
				\label{H_inv1}
				H_{abcd} &:=& \frac{1}{2} \left ( \uacc{h}_{ac} \uacc{h}_{bd} + \uacc{h}_{ad} \uacc{h}_{bc}  - \dfrac{2}{n-2}\uacc{h}_{ab} \uacc{h}_{cd} \right). 
			\end{eqnarray}
The relation~\eqref{Q2pi} is a map from the variables $(\tilde{\pi}^{ab}, \tilde{\Pi}^{IJ}) \longmapsto (Q_{aI})$. The inverse map $(Q_{aI}) \longmapsto (\tilde{\pi}^{ab}, \tilde{\Pi}^{IJ})$ is given by~\eqref{defADMpfinal} and
	\begin{eqnarray}\label{Pi=G}
		\tilde{\Pi}^{IJ} = 2 \tilde{\Pi}^{a[I} Q_{a}{}^{J]}.
	\end{eqnarray}    
	Therefore, $\tilde{\Pi}^{IJ}$ has the same mathematical expression as $\tilde{\mathcal{G}}^{IJ}$ given in~\eqref{Gauss1}. Thus, the parametrization for $Q_{aI}$ given by~\eqref{Q2pi} allows us to naturally handle the Gauss constraint. 
	
	Note that we can rephrase the previous deduction of the variables involved: $q_{ab}$, $\tilde{\pi}^{ab}$, and $\tilde{\Pi}^{IJ}$. This alternative viewpoint is also illustrative and might be preferred by some readers. The new perspective is as follows: We could have started the analysis by defining the variables $\tilde{\Pi}^{IJ}$ through the relation~\eqref{Pi=G}, keeping in mind that the symplectic reduction needs to take into account the Gauss constraint at some point of the analysis. Again, the relation~\eqref{Pi=G} can be thought as  $n (n-1)/2$ equations for $n (n-1)$ unknowns $Q_{aI}$, and its solution is precisely~\eqref{Q2pi} where the variables $\tilde{\pi}^{ab}$ now play the role of the independent
	free variables. Due to the fact the map is invertible, then we compute the inverse map and get~\eqref{defADMpfinal}. So, the interpretation changes, but the formulas are the same.

	\subsection{Symplectic reduction}\label{reduction2}
	If we substitute the new parametrization for $Q_{aI}$ given by~\eqref{Q2pi}, then the sympletic structure of the Hamiltonian action~\eqref{S_QP} becomes
	\begin{eqnarray}\label{ss}
		2 \tilde{\Pi}^{aI} \dot{Q}_{aI} &=& \tilde{\pi}^{ab} \dot{q}_{ab} - \dfrac{2}{n-2} \partial_{t} \left( \tilde{\pi}^{ab} q_{ab} \right) \nonumber\\
		&& -  \uacc{h}_{ab}  \tilde{\Pi}^{b}{}_{J} \left(\eta_{IK} + \sigma n_{I} n_{K} \right) \tilde{\Pi}^{JK} \partial_{t} \tilde{\Pi}^{aI}. \quad
	\end{eqnarray}
	On the other hand, the first-class constraints given in~\eqref{S_QP} acquire the form
	\begin{subequations}
		\begin{eqnarray}
			\label{G_Y}
			\tilde{\mathcal{G}}^{IJ} &=& \tilde{\Pi}^{IJ}, \\
			\label{D_Y}
			\tilde{\mathcal{D}}_{a} &=& - q_{ab} D_{c} \tilde{\pi}^{bc}  - \dfrac{1}{2} \Gamma_{aIJ} \tilde{\Pi}^{IJ} \notag \\
			&& - \dfrac{1}{2q} q_{ab} \tilde{\Pi}^{bI} \tilde{\Pi}^{cJ} \nabla_{c} \tilde{\Pi}_{IJ},	\\
			\label{H_Y}
			\tilde{\tilde{\mathcal{H}}} &=& - \sigma qR  -Q_{abcd}\tilde{\pi}^{ab} \tilde{\pi}^{cd} + 2 \sigma q \Lambda \notag \\
			&&+ \dfrac{1}{4} \tilde{\Pi}^{IJ} \left( \tilde{\Pi}_{IJ} + 2 \sigma n_{I} n^{K} \tilde{\Pi}_{JK} \right),
		\end{eqnarray}
	\end{subequations}
	where 
	\begin{eqnarray}
        \label{Q_inv}
        Q_{abcd} :=\frac{1}{2} \left ( q_{ac} q_{bd} + q_{ad} q_{bc} - \dfrac{2}{n-2} q_{ab} q_{cd} \right ),
    \end{eqnarray}
    $q=\det{(q_{ab})}$, $R := q^{ac} q^{bd} R_{abcd}$ is the scalar curvature of the Levi-Civita connection $\Gamma^a{}_{bc}$ defined by the metric $q_{ab}$ on $\Sigma$, $D_a$ is the covariant derivative with respect to $\Gamma^a{}_{bc}$, and so
	\begin{eqnarray}
	    \label{Dp}
		D_{a} \tilde{\pi}^{bc} &:=& \partial_{a} \tilde{\pi}^{bc}  + \Gamma^{b}{}_{ad} \tilde{\pi}^{dc} + \Gamma^{c}{}_{ad} \tilde{\pi}^{bd} - \Gamma^{d}{}_{ad} \tilde{\pi}^{bc}, \quad
	\end{eqnarray}
	whereas $\nabla_{a} \tilde{\Pi}^{IJ} = \partial_{a} \tilde{\Pi}^{IJ} - \Gamma^{b}{}_{ab} \tilde{\Pi}^{IJ} + \Gamma_{a}{}^{I}{}_{K} \tilde{\Pi}^{KJ} + \Gamma_{a}{}^{J}{}_{K} \tilde{\Pi}^{IK}$. In order to obtain~\eqref{G_Y},~\eqref{D_Y}, and~\eqref{H_Y}, we used $\uacc{h}_{ab} \tilde{\Pi}^{a}{}_{I} \tilde{\Pi}^{b}{}_{J} = \eta_{IJ} - \sigma n_{I} n_{J}$ and $\tilde{\Pi}^{aI} \tilde{\Pi}^{bJ} R_{abIJ} = qR$, where the latter expression is easily derived by computing $\nabla_{[a} \nabla_{b]} \tilde{\Pi}^{cI}= 0$ and using the standard definition for the scalar curvature $R$. In this way, the remaining task is to handle the terms involving the Gauss constraint in the constraints~\eqref{D_Y} and~\eqref{H_Y}.
	
	Factoring out the Gauss constraint, which requires to integrate by parts the last term of~\eqref{D_Y} and to neglect the resulting boundary term, the action~\eqref{S_QP} becomes after redefining the Lagrange multiplier of the Gauss constraint and neglecting the boundary term of~\eqref{ss}
	\begin{eqnarray}
		\label{S_ADM_G}
		S &= & \int_{\mathbb{R} \times \Sigma}\!\!\! dt d^{n-1}x\Big (  \tilde{\pi}^{ab} \dot{q}_{ab} - 2 N^{a}\tilde{\mathcal{C}}_{a} - N \tilde{\mathcal{C}} -\Lambda_{IJ} \tilde{\mathcal{G}}^{I J} \Big ), \notag \\
	\end{eqnarray}
	where $N:= \sqrt{q} \uac{N}$, $\tilde{\mathcal{G}}^{IJ}$ is given by~\eqref{G_Y} and the diffeomorphism and Hamiltonian constraints are given by
	\begin{subequations}
		\begin{eqnarray}
			\label{Ca}
			\tilde{\mathcal{C}}_{a} &:=& - q_{ab} D_{c} \tilde{\pi}^{bc}, \\
			\label{C}
			\tilde{\mathcal{C}} &:=& - \sigma \sqrt{q} R  - \dfrac{1}{\sqrt{q}} Q_{abcd}\tilde{\pi}^{ab} \tilde{\pi}^{cd} + 2 \sigma \sqrt{q} \Lambda.
		\end{eqnarray}
	\end{subequations}
	Therefore, on the hypersurface where the Gauss constraint is satisfied, $\tilde{\mathcal{G}}^{IJ}=0$, we automatically get the ADM formulation of general relativity described by the canonical phase-space variables $(q_{ab}, \tilde{\pi}^{ab})$. However, from Dirac's approach to constrained systems~\cite{DiracBook}, we must impose in~\eqref{S_ADM_G} the gauge condition that fixes the gauge freedom generated by the Gauss constraint. Therefore, we consider the variable $\tilde{\Pi}^{IJ}$ as momentum variable and define $Y_{IJ} = - Y_{JI}$ as its corresponding configuration variable. This means that we must add the term $\tilde{\Pi}^{IJ} {\dot Y}_{IJ}$ and also to impose the constraint $Y_{IJ} \approx 0$. Hence, the action~\eqref{S_ADM_G} becomes
	\begin{eqnarray}
		\label{S_ADM_GP}
		S &= & \int_{\mathbb{R} \times \Sigma} \! dt d^{n-1}x\Big ( \tilde{\pi}^{ab} \dot{q}_{ab} + \tilde{\Pi}^{IJ}\dot{Y}_{IJ}   - 2 N^{a} \tilde{\mathcal{C}}_{a} - N \tilde{\mathcal{C}}  \notag \\
		&& -\Lambda_{IJ} \tilde{\mathcal{G}}^{IJ}  - \tilde{\rho}^{IJ} Y_{IJ} \Big), 
	\end{eqnarray}
	where $\tilde{\rho}^{IJ} = - \tilde{\rho}^{JI}$ is the Lagrange multiplier that imposes the constraint $Y_{IJ} \approx 0$. Clearly $\tilde{\mathcal{G}}^{IJ} \approx 0$ and $Y_{IJ} \approx 0$ are second-class constraints (with a possible modification of $\tilde{\mathcal{C}}_{a}$ and $\tilde{\mathcal{C}} $ involving the second-class constraints). Making them strongly equal to zero, we get the ADM formulation of general relativity in $n$ dimensions given by the Hamiltonian action principle~\cite{ThieBook} (see, of course, the original work of Arnowitt, Desser, and Misner in four dimensions~\cite{ADM})
	\begin{equation}\label{getting_ADM}
		S =  \int_{\mathbb{R} \times \Sigma}dt d^{n-1}x\Big (  \tilde{\pi}^{ab} \dot{q}_{ab}  - 2 N^{a}\tilde{\mathcal{C}}_{a} - N \tilde{\mathcal{C}} \Big),
	\end{equation}
	where the phase space is described by the canonical pair $(q_{ab}, \tilde{\pi}^{ab})$ and the constraints are given by \eqref{Ca} and \eqref{C}.

	\section{From ADM to $SO(n-1,1)$ [$SO(n)$] phase-space variables}\label{ADM_to_Palatini}
	
	Now, we start from the Hamiltonian formulation of general relativity in terms of the ADM variables $(q_{ab},\tilde{\pi}^{ab})$, given by the action~\cite{ThieBook}
	\begin{eqnarray}
		\label{S_ADM}
		S =  \int_{\mathbb{R} \times \Sigma}dt d^{n-1}x\Big (  \tilde{\pi}^{ab} \dot{q}_{ab}  - 2 N^{a}\tilde{\mathcal{C}}_{a} - N \tilde{\mathcal{C}} \Big),
	\end{eqnarray}
	with the first-class constraints
	\begin{subequations}
		\begin{eqnarray}
			\label{Ca_ADM}
			\tilde{\mathcal{C}}_{a} &:=& - q_{ab} D_{c} \tilde{\pi}^{bc}, \\
			\label{C_ADM}
			\tilde{\mathcal{C}} &:=& - \sigma \sqrt{q} R  - \dfrac{1}{\sqrt{q}} Q_{abcd}\tilde{\pi}^{ab} \tilde{\pi}^{cd} + 2 \sigma \sqrt{q} \Lambda,
		\end{eqnarray}
	\end{subequations}
	where $q=\det(q_{ab})$, $R$ is the scalar curvature of the $(n-1)$-dimensional hypersurface $\Sigma$, $\Lambda$ is the cosmological constant, and $Q_{abcd}$ is defined in~\eqref{Q_inv}. Also, $D_{a}$ is the covariant derivative compatible spatial metric $q_{ab}$, i.e., 
	\begin{equation}
		\label{Dq}
		D_{a} q_{bc} := \partial_{a} q_{bc} - \Gamma^{d}{}_{ab} q_{dc} - \Gamma^{d}{}_{ac} q_{bd} =0.
	\end{equation}

	Next, the idea is to perform a ``lifting'' of the ADM Hamiltonian formulation to the Hamiltonian formulation~\eqref{S_QP} where the local $SO(n-1,1)$ [$SO(n)$] symmetry is manifest. This is relevant because we can directly get such a formulation without starting from the $n$-dimensional Palatini action, as it is made in Refs.~\cite{Montesinos2001,Bodendorfer1301a,Bodendorfer1301b}.
	
	This bottom-up approach might also be useful, for instance, for researchers familiar with Hamiltonian formulations of metric theories---alternative to Einstein's general relativity---that desire to see how they look in the first-order formalism without redoing the computations from the corresponding first-order Lagrangians of the metric formulations.
	
	To begin with, we introduce the densitized vector $\tilde{\Pi}^{aI}$, where the internal indices $I, J, K, \ldots$ that take on the values  $0,1,\ldots, n-1$ are lowered or raised with the $n$-dimensional Minkowski ($\sigma=-1$) or Euclidean ($\sigma=1$) metric $(\eta_{IJ}):= \mbox{diag}(\sigma, 1, \ldots, 1)$. We now parametrize the inverse of the spatial metric $q_{ab}$ as a function of the variables $\tilde{\Pi}^{aI}$ as
	\begin{equation}
		\label{q2Pi}
		q^{ab} = h^{-1/(n-2)} \eta_{IJ} \tilde{\Pi}^{aI} \tilde{\Pi}^{bJ},
	\end{equation}
	where $h:= \det(\eta_{IJ} \tilde{\Pi}^{aI} \tilde{\Pi}^{bJ})$ is a tensor density of weight $2(n-2)$. By hypothesis, $q= \det{(q_{ab})} >0$, and because of $q= \det(q_{ab}) = h^{1/(n-2)}$, then $h>0$.  
	
	Similarly, we introduce the variable $Q_{aI}$, and parametrize the ADM momentum as
	\begin{equation}
		\label{pi_sol}
		\tilde{\pi}^{ab} =  h^{-1/(n-2)} \left(H^{-1}\right)^{abcd} \uacc{h}_{e(c} \tilde{\Pi}^{eI} Q_{d)I},
	\end{equation}
	where
	\begin{eqnarray}
			\label{H_def}
			\left(H^{-1}\right)^{abcd} &:=&  \frac{1}{2} \left ( \tilde{\tilde{h}}^{ac} \tilde{\tilde{h}}^{bd} + \tilde{\tilde{h}}^{ad} \tilde{\tilde{h}}^{bc}  - 2 \tilde{\tilde{h}}^{ab} \tilde{\tilde{h}}^{cd}  \right )
	\end{eqnarray}
	is a tensor density of weight $4$ and $\uacc{h}_{ab}$ is the densitized metric inverse of $\tilde{\tilde{h}}^{ab} := \tilde{\Pi}^{aI} \tilde{\Pi}^{b}{}_{I}$. Note that the definition of $\tilde{\pi}^{ab}$ is exactly the same expression given in~\eqref{defADMpfinal}. Moreover, notice that $\left(H^{-1}\right)^{abcd}$ and $H_{abcd}$, defined in~\eqref{H_inv1}, fulfill the relation $H_{abef} \left( H^{-1}\right)^{efcd} = (1/2) \left( \delta^{c}_{a} \delta^{d}_{b} + \delta^{d}_{a} \delta^{c}_{b} \right)$.
	
	Thus, the relations~\eqref{q2Pi} and~\eqref{pi_sol} are invariant under $SO(n-1,1)$ [$SO(n)$] transformations $\tilde{\Pi}^{aI} \mapsto \Lambda^{I}{}_{J} \tilde{\Pi}^{aJ}$ when $\sigma=-1$ [$\sigma=1$]. 
	
	In this way, the initial ADM variables $(q_{ab}, \tilde{\pi}^{ab})$ of the Hamiltonian formulation~\eqref{S_ADM} have been parametrized in terms of the variables $(Q_{aI}, \tilde{\Pi}^{aI})$. The variables $(Q_{aI}, \tilde{\Pi}^{aI})$ are not independent and we must impose the Gauss constraint
	\begin{equation}
		\label{Gauss_def}
		\tilde{\mathcal{G}}^{IJ} := 2 \tilde{\Pi}^{a[I} Q_{a}{}^{J]} \approx 0
	\end{equation}
among them, which also generates their local $SO(n-1,1)$ [$SO(n)$] transformations.
	
Using~\eqref{q2Pi} and~\eqref{pi_sol}, the symplectic structure in~\eqref{S_ADM} acquires the form
	\begin{eqnarray}
		\tilde{\pi}^{ab} \dot{q}_{ab}  &=& -2 Q_{aI} \partial_{t}\tilde{\Pi}^{aI} +  \uacc{h}_{ab}  \tilde{\Pi}^{b}{}_{J} \tilde{\mathcal{G}}^{JK} \big(\eta_{IK} \notag \\
			&&  + \sigma n_{I} n_{K} \big)  \partial_{t}\tilde{\Pi}^{aI} \notag  \\
			\label{ss2}
			&=& 2 \tilde{\Pi}^{aI} \dot{Q}_{aI} -2 \partial_{t}\left( \tilde{\Pi}^{aI} Q_{aI} 
 \right) \notag \\
			&& +  \uacc{h}_{ab}  \tilde{\Pi}^{b}{}_{J} \big(\eta_{IK}  + \sigma n_{I} n_{K} \big) \tilde{\mathcal{G}}^{JK} \partial_{t}\tilde{\Pi}^{aI}.
	\end{eqnarray}
	Therefore, $(Q_{aI}, \tilde{\Pi}^{aI})$ are indeed canonical variables up to the Gauss constraint~\eqref{Gauss_def}. The remaining task is to rewrite the first-class constraints~\eqref{Ca_ADM} and \eqref{C_ADM} in terms of $(Q_{aI}, \tilde{\Pi}^{aI})$. To do that, we introduce the covariant derivative $\nabla_{a}$ compatible with $\tilde{\Pi}^{aI}$ as
	\begin{equation}
		\nabla_{a}\tilde{\Pi}^{b I} :=  \partial_{a}\tilde{\Pi}^{b I} + \Gamma^{b}{}_{a c}\tilde{\Pi}^{c I}  - \Gamma^{c}{}_{a c}\tilde{\Pi}^{b I} + \Gamma_{a}{}^{I}{}_{J}\tilde{\Pi}^{b J} = 0,
	\end{equation}
	with $\Gamma^c{}_{ab}= \Gamma^c{}_{ba}$ and $\Gamma_{aIJ}=-\Gamma_{aJI}$. Therefore, using~\eqref{q2Pi},~\eqref{pi_sol}, and the identities $\uacc{h}_{ab} \tilde{\Pi}^{a}{}_{I} \tilde{\Pi}^{b}{}_{J} = \eta_{IJ} - \sigma n_{I} n_{J}$ and $\tilde{\Pi}^{aI} \tilde{\Pi}^{bJ} R_{abIJ} = qR$, the constraints~\eqref{Ca_ADM} and \eqref{C_ADM} become
	\begin{subequations}
		\begin{eqnarray}
			\tilde{\mathcal{C}}_{a} &=& 2\tilde{\Pi}^{bI} \partial_{[a} Q_{b]I} - Q_{aI} \partial_{b}\tilde{\Pi}^{bI} + \dfrac{1}{2} \Gamma_{aIJ} \tilde{\mathcal{G}}^{IJ} \notag \\
			&& + \dfrac{1}{2} \uacc{h}_{ab} \tilde{\Pi}^{bI} \tilde{\Pi}^{cJ} \nabla_{c} \tilde{\mathcal{G}}_{IJ},	\\
			\tilde{\mathcal{C}} &=&  h^{-\frac{1}{2(n-2)}} \bigg[ - \sigma \tilde{\Pi}^{a I} \tilde{\Pi}^{b J}R_{abIJ} +  2 \tilde{\Pi}^{a[I}\tilde{\Pi}^{|b|J]}Q_{a I}Q_{b J} \notag \\
            && \!\!\! - \dfrac{1}{4} \tilde{\mathcal{G}}^{IJ} \big( \tilde{\mathcal{G}}_{IJ} +  2 \sigma n_{I} n^{K} \tilde{\mathcal{G}}_{JK}  \big) \bigg]  + 2 \sigma  h^{\frac{1}{2(n-2)}} \Lambda, 
		\end{eqnarray}
	\end{subequations}
	where we used $\nabla_{a} \tilde{\mathcal{G}}^{IJ} = \partial_{a} \tilde{\mathcal{G}}^{IJ} - \Gamma^{b}{}_{ab} \tilde{\mathcal{G}}^{IJ} + \Gamma_{a}{}^{I}{}_{K} \tilde{\mathcal{G}}^{KJ} + \Gamma_{a}{}^{J}{}_{K} \tilde{\mathcal{G}}^{IK}$. Therefore, adding the Gauss constraint~\eqref{Gauss_def} to the action~\eqref{S_ADM} and collecting all the terms proportional to this constraint, the action~\eqref{S_ADM} acquires the form
		\begin{eqnarray}
		\label{S_ADM_final}
		S &= & \int_{\mathbb{R} \times \Sigma} dt d^{n-1}x  \Big( 2\tilde{\Pi}^{a I}\dot{Q}_{a I} -\lambda_{IJ} \tilde{\mathcal{G}}^{I J}  \notag \\
			&& - 2 N^{a}\tilde{\mathcal{D}}_{a} - \underaccent{\tilde}{N} \tilde{\tilde{\mathcal{H}}} \Big), 
	\end{eqnarray}
	with $\uac{N} = h^{-\frac{1}{2(n-2)}} N$ and the first-class constraints are given by
	\begin{subequations}
		\begin{eqnarray}
			\label{G_ADM_final}
			\tilde{\mathcal{G}}^{IJ} &=& 2 \tilde{\Pi}^{a[I} Q_{a}{}^{J]}, \\
			\tilde{\mathcal{D}}_{a} &:=& 2 \tilde{\Pi}^{bI} \partial_{[a} Q_{b]I} - Q_{aI} \partial_{b}\tilde{\Pi}^{bI}, \\
			\label{H_ADM_final}
			\tilde{\tilde{\mathcal{H}}} &:=& - \sigma \tilde{\Pi}^{a I} \tilde{\Pi}^{b J}R_{abIJ} +  2 \tilde{\Pi}^{a[I}\tilde{\Pi}^{|b|J]}Q_{a I}Q_{b J} \notag \\
			&& + 2 \sigma  h^{1/(n-2)} \Lambda,
		\end{eqnarray}
	\end{subequations}
where we neglected the boundary term in~\eqref{ss2} and the term that comes out from the integration by parts of $\nabla_{a} \tilde{\mathcal{G}}^{IJ}$, and also $\lambda_{IJ}= -\lambda_{JI}$.

Note that the formulation encompassed by the action~\eqref{S_ADM_final} with the constraints~\eqref{G_ADM_final}--\eqref{H_ADM_final} is precisely the Hamiltonian formulation of the the $n$-dimensional Palatini action~\eqref{S_QP}. 

Let us make a remark. In a spacetime of dimension four, we can also perform the bottom-up approach of Sec.~\ref{ADM_to_Palatini} to go from the ADM formulation to the $SO(3,1)$ [$SO(4)$ if $\sigma=1$] Hamiltonian formulation of the Holst action~\cite{Holst9605}, featuring only first-class constraints, given in Refs.~\cite{Montesinos1801,Montesinos2004a}. This is achieved by using~\eqref{pi_sol} and the relationship between the configuration variable $Q_{aI}$ and the configuration variables $C_{aI}$ or $K_{aI}$ described in Refs.~\cite{Montesinos1801,Montesinos2004a}. Similarly, the top-down approach to go from the $SO(3,1)$ [$SO(4)$] Hamiltonian formulation~\cite{Montesinos1801,Montesinos2004a} of the Holst action to the ADM formulation can also be made following the approach of Sec.~\ref{Lorentz_to_ADM} but involving the configuration variables $C_{aI}$ or $K_{aI}$.

	\section{From $SO(n-1)$ to ADM variables and vice versa}\label{vectors_ADM}
	Imposing the time gauge in the formulation given by the action~\eqref{S_QP} results in \cite{Montesinos2001}
	\begin{eqnarray}
		\label{S_QP_tg}
		S &= & \int_{\mathbb{R} \times \Sigma} dt d^{n-1}x \Big( 2\tilde{\Pi}^{ai}\dot{Q}_{ai} -\lambda_{ij} \tilde{\mathcal{G}}^{ij} \notag \\
		&& - 2 N^{a}\tilde{\mathcal{D}}_{a} - \underaccent{\tilde}{N} \tilde{\tilde{\mathcal{H}}} \Big),
	\end{eqnarray}
	with the first-class constraints given by
	\begin{subequations}
		\begin{eqnarray}
			\label{G_tg}
			\tilde{\mathcal{G}}^{ij} &=& 2 \tilde{\Pi}^{a[i} Q_{a}{}^{j]}, \\
			\tilde{\mathcal{D}}_{a} &=& 2\tilde{\Pi}^{bi} \partial_{[a} Q_{b]i} - Q_{ai} \partial_{b}\tilde{\Pi}^{bi}, \\
			\label{H_tg}
			\tilde{\tilde{\mathcal{H}}} &=& - \sigma \tilde{\Pi}^{ai} \tilde{\Pi}^{bj}R_{abij} +  2 \tilde{\Pi}^{a[i}\tilde{\Pi}^{|b|j]}Q_{ai}Q_{bj} \notag \\
			&& + 2 \sigma  |\det(\tilde{\Pi}^{ai})|^{2/(n-2)} \Lambda.
		\end{eqnarray}
	\end{subequations}
	The indices $i, j, k, \ldots$ taking on the values $1,2,\ldots, n-1$ are $SO(n-1)$ valued, which are raised or lowered with the ($n-1$)-dimensional Euclidean metric $\delta_{ij}$. Here, $Q_{ai}$ and $\tilde{\Pi}^{ai}$ are canonical variables that transform as $SO(n-1)$ vectors under local $SO(n-1)$ rotations. Geometrically, $Q_{ai}$ is related to the extrinsic curvature and $\tilde{\Pi}^{ai}$ is the densitized triad defined on $\Sigma$. Also, $\det(\tilde{\Pi}^{ai})$ is a tensor density of weight $n-2$. On the other hand, the Gauss constraint $\tilde{\mathcal{G}}^{ij}$ generates the local $SO(n-1)$ rotations whereas the diffeomorphism $\tilde{\mathcal{D}}_{a}$ and scalar $\tilde{\tilde{\mathcal{H}}}$ constraints are still responsible of generating the spacetime diffeomorphisms. 
	
	In the next two subsections we show how to go from this Hamiltonian formulation to the ADM one (top-down approach) and vice versa (bottom-up approach).
	
	\subsection{From $SO(n-1)$ to ADM variables}
	\label{SO_to_ADM}
	The path we follow is analogous to the one described in Sec.~\ref{Lorentz_to_ADM}; hence, we begin with the formulation given by the action~\eqref{S_QP_tg} and the first-class constraints \eqref{G_tg}--\eqref{H_tg}. First, let $\uac{\Pi}_{ai}$ be the inverse of $\tilde{\Pi}^{ai}$, so $\uac{\Pi}_{ai} \tilde{\Pi}^{aj}=\delta_{i}^{j}$ and $\uac{\Pi}_{ai} \tilde{\Pi}^{bi}=\delta_{a}^{b}$. Then, we define the spatial metric $q_{ab}$ and its inverse $q^{ab}$ as
	\begin{subequations}
		\begin{eqnarray}
			\label{q_def}
			q_{ab} &:=& \mid \det(\tilde{\Pi}^{ai})\mid^{2/(n-2)} \uac{\Pi}_{ai} \uac{\Pi}_{b}{}^{i}, \\
			\label{q_inv_def}
			q^{ab} &=& \mid \det(\tilde{\Pi}^{ai}) \mid^{-2/(n-2)} \tilde{\Pi}^{ai} \tilde{\Pi}^{b}{}_{i}.
		\end{eqnarray}
	\end{subequations}
Using these expressions, we define the quantities
	\begin{subequations}
		\begin{eqnarray}
			\label{Q_SO3}
			\left( Q^{-1} \right)^{abcd} &:=& \frac{1}{2} \left ( q^{ac} q^{bd} + q^{ad} q^{bc}  - 2 q^{ab} q^{cd} \right ),\\
			\label{Q_inv_SO3}
			Q_{abcd} &=& \frac{1}{2} \left( q_{ac} q_{bd} + q_{ad} q_{bc} - \dfrac{2}{n-2} q_{ab} q_{cd}\right), \quad \quad \;
		\end{eqnarray}
	\end{subequations}
    which satisfy $Q_{abef} \left( Q^{-1}\right)^{efcd} = (1/2) \left( \delta^{c}_{a} \delta^{d}_{b} + \delta^{d}_{a} \delta^{c}_{b} \right)$.
 
	On the other hand, in the ADM formulation, the canonical variables are the spatial metric $q_{ab}$ and its conjugated momentum $\tilde{\pi}^{ab}$. So, the remaining task is to define the ADM momentum $\tilde{\pi}^{ab}$ in terms of the $SO(n-1)$ variables:
	\begin{eqnarray}
		\label{pi_def2}
		\tilde{\pi}^{ab} &:=&  |\det(\tilde{\Pi}^{ai})|^{2/(n-2)} \left( Q^{-1} \right)^{abcd} \uac{\Pi}_{(c}{}^{i} Q_{d)i}.
	\end{eqnarray}
	Hence, to reach the ADM formulation we use~\eqref{pi_def2} and solve for $Q_{ai}$. We recall that $Q_{ai}$ are $(n-1)^{2}$ variables. Thus, in order to solve~\eqref{pi_def2}, which has $n(n-1)/2$ equations, we must introduce $(n-1)(n-2)/2$ independent variables $\tilde{\Pi}^{ij} =-\tilde{\Pi}^{ji}$. Therefore, the solution of~\eqref{pi_def2} is
	\begin{eqnarray}
		\label{Q2pi_tg}
		Q_{ai} &=&  |\det(\tilde{\Pi}^{ai})|^{-2/(n-2)} Q_{abcd} \tilde{\Pi}^{b}{}_{i} \tilde{\pi}^{cd}  + \dfrac{1}{2} \uac{\Pi}_{aj} \tilde{\Pi}^{j}{}_{i}. \quad\quad
	\end{eqnarray}
	
	By using~\eqref{Q2pi_tg}, we rewrite the symplectic structure given in~\eqref{S_QP_tg} and we get
	\begin{eqnarray}\label{ss_tg}
		2 \tilde{\Pi}^{ai} \dot{Q}_{ai} &=& \tilde{\pi}^{ab} \dot{q}_{ab} - \dfrac{2}{n-2} \partial_{t} \left( \tilde{\pi}^{ab} q_{ab} \right) \nonumber\\
		&& +   \uac{\Pi}_{a}{}^{j} \tilde{\Pi}_{ij} \partial_{t}\tilde{\Pi}^{ai},
	\end{eqnarray}
	and we also rewrite the first-class constraints given in~\eqref{S_QP_tg} and we obtain
	\begin{subequations}
		\begin{eqnarray}
			\label{G_tg2}
			\tilde{\mathcal{G}}^{ij} &=& \tilde{\Pi}^{ij}, \\
			\label{D_tg2}
			\tilde{\mathcal{D}}_{a} &=& - q_{ab} D_{c} \tilde{\pi}^{bc}  - \dfrac{1}{2} \Gamma_{aij} \tilde{\Pi}^{ij} \notag \\
			&& - \dfrac{1}{2} \uac{\Pi}_{a}{}^{i} \tilde{\Pi}^{bj} \nabla_{b} \tilde{\Pi}_{ij},	\\
			\label{H_tg2}
			\tilde{\tilde{\mathcal{H}}} &=& - \sigma qR  - Q_{abcd}\tilde{\pi}^{ab} \tilde{\pi}^{cd} + 2 \sigma q \Lambda \notag \\
			&&+ \dfrac{1}{4} \tilde{\Pi}^{ij} \tilde{\Pi}_{ij},
		\end{eqnarray}
	\end{subequations}
	where $q=\det{(q_{ab})}$, $R := q^{ac} q^{bd} R_{abcd}$ is the scalar curvature of the Levi-Civita connection $\Gamma^{a}{}_{bc}$, and $D_{a}$ is the covariant derivative with respect to the Levi-Civita connection [see~\eqref{Dp}].  We also used the identity $\tilde{\Pi}^{ai} \tilde{\Pi}^{bj} R_{abij} = qR$.
	
	By substituting~\eqref{ss_tg},~\eqref{G_tg2},~\eqref{D_tg2}, and~\eqref{H_tg2} into the action~\eqref{S_QP_tg} and factoring out all the terms proportional to the Gauss constraint leads us to the action
	\begin{eqnarray}
		\label{S_ADM_G_tg}
		S &= & \int_{\mathbb{R} \times \Sigma}dt d^{n-1}x\Big (  \tilde{\pi}^{ab} \dot{q}_{ab} - 2 N^{a}\tilde{\mathcal{C}}_{a} - N \tilde{\mathcal{C}} -\Lambda_{ij} \tilde{\mathcal{G}}^{ij} \Big ), \notag \\
	\end{eqnarray}
	where $N = \sqrt{q} \uac{N}$, $\tilde{\mathcal{G}}^{ij}$ is given by~\eqref{G_tg2}, and the diffeomorphism and Hamiltonian constraints are given by
	\begin{subequations}
		\begin{eqnarray}
			\label{Ca2}
			\tilde{\mathcal{C}}_{a} &:=& - q_{ab} D_{c} \tilde{\pi}^{bc}, \\
			\label{C2}
			\tilde{\mathcal{C}} &:=& - \sigma \sqrt{q} R  - \dfrac{1}{\sqrt{q}} Q_{abcd}\tilde{\pi}^{ab} \tilde{\pi}^{cd} + 2 \sigma \sqrt{q} \Lambda.
		\end{eqnarray}
	\end{subequations}
	Thus, just as in Sec.~\ref{Lorentz_to_ADM}, we obtain the ADM formulation in the hypersurface where the Gauss constraint $\tilde{\mathcal{G}}^{ij}=0$ is satisfied. Using Dirac's approach to constrained systems~\cite{DiracBook}, we introduce the variable $Y_{ij} = - Y_{ji}$ as the configuration variable conjugated to $\tilde{\Pi}^{ij}$. Therefore, we add the term $\tilde{\Pi}^{ij} \dot{Y}_{ij}$ and the constraint $Y_{ij} \approx 0$. Hence, the action~\eqref{S_ADM_G_tg} becomes
	\begin{eqnarray}
		\label{S_ADM_GP2}
		S &= & \int_{\mathbb{R} \times \Sigma}dt d^{n-1}x\Big ( \tilde{\pi}^{ab} \dot{q}_{ab} + \tilde{\Pi}^{ij} \dot{Y}_{ij} - 2 N^{a}\tilde{\mathcal{C}}_{a} - N \tilde{\mathcal{C}}  \notag \\
		&& -\Lambda_{ij} \tilde{\mathcal{G}}^{ij}  - \tilde{\rho}^{ij} Y_{ij} \Big), 
	\end{eqnarray}
	where $\tilde{\rho}^{ij} = - \tilde{\rho}^{ji}$ is the Lagrange multiplier that imposes the constraint $Y_{ij} \approx 0$. Since $\tilde{\mathcal{G}}^{ij} \approx 0$ and $Y_{ij} \approx 0$ are second-class constraints (with a possible modification of $\tilde{\mathcal{C}}_{a}$ and $\tilde{\mathcal{C}} $ involving the second-class constraints), we can make them strongly equal to zero, which results in the ADM formulation of general relativity given by the Hamiltonian action principle~\cite{ThieBook}
	\begin{equation}
	    \label{S_ADM2}
		S =  \int_{\mathbb{R} \times \Sigma}dt d^{n-1}x\Big (  \tilde{\pi}^{ab} \dot{q}_{ab}  - 2 N^{a}\tilde{\mathcal{C}}_{a} - N \tilde{\mathcal{C}} \Big),
	\end{equation}
	where $(q_{ab}, \tilde{\pi}^{ab})$ are the canonical variables and the constraints are given by \eqref{Ca2} and \eqref{C2}.

\subsection{From ADM to $SO(n-1)$ variables}
	\label{ADM_to_SO}
	Similarly to what is made in Sec.~\ref{ADM_to_Palatini}, we can make a ``lifting'' from the ADM formulation given in~\eqref{S_ADM2} to the $SO(n-1)$ formulation of~\eqref{S_QP_tg}. To do so, we have to enlarge the phase space to account for the additional gauge freedom. Thus, we must replace the initial ADM variables $(q_{ab}, \tilde{\pi}^{ab})$ with the variables $(Q_{ai}, \tilde{\Pi}^{ai})$, where the indices $i,j,k,\ldots$ take on the values $1,\ldots, n-1$ and are raised or lowered with the ($n-1$)-dimensional Euclidean metric $\delta_{ij}$. The enlargement of the phase space requires us to impose an additional constraint among the variables $(Q_{ai}, \tilde{\Pi}^{ai})$ because they are not independent among themselves. Hence, we must add the Gauss constraint
	\begin{equation}
		\label{Gauss_def2}
		\tilde{\mathcal{G}}^{ij} := 2 \tilde{\Pi}^{a[i} Q_{a}{}^{j]} \approx 0,
	\end{equation}
	which generates the local $SO(n-1)$ rotations.
	
	Thus, to perform the ``lifting'' from the ADM variables we use~\eqref{q_def} and~\eqref{pi_def2}, where for $Q^{abcd}$ in~\eqref{pi_def2} we use~\eqref{q_inv_def}. Then, the symplectic structure of~\eqref{S_ADM2} becomes
	\begin{eqnarray}
		\label{ss_ADM2}
		\tilde{\pi}^{ab} \dot{q}_{ab}  &=& 2 \tilde{\Pi}^{ai} \dot{Q}_{ai} - 2 \partial_{t}\left( \tilde{\Pi}^{ai} Q_{ai} \right) \notag \\
		&& -   \uac{\Pi}_{a}{}^{j} \tilde{\mathcal{G}}_{ij} \partial_{t}\tilde{\Pi}^{ai}.
	\end{eqnarray}
	Therefore, $(Q_{ai}, \tilde{\Pi}^{ai})$ are indeed canonical variables up to the Gauss constraint~\eqref{Gauss_def2}. On the other hand, using~\eqref{q_def} and~\eqref{pi_def2}, the constraints~\eqref{Ca2} and \eqref{C2} become
	\begin{subequations}
		\begin{eqnarray}
			\tilde{\mathcal{C}}_{a} &=& 2\tilde{\Pi}^{bi} \partial_{[a} Q_{b]i} - Q_{ai} \partial_{b}\tilde{\Pi}^{bi} + \dfrac{1}{2} \Gamma_{aij} \tilde{\mathcal{G}}^{ij} \notag \\
			&& + \dfrac{1}{2} \uac{\Pi}_{a}{}^{i} \tilde{\Pi}^{bj} \nabla_{b} \tilde{\mathcal{G}}_{ij},	\\
			\tilde{\mathcal{C}} &=& |\det(\tilde{\Pi}^{ai})|^{-1/(n-2)} \bigg( - \sigma \tilde{\Pi}^{ai} \tilde{\Pi}^{bj} R_{abij} \notag \\
            && +  2 \tilde{\Pi}^{a[i}\tilde{\Pi}^{|b|j]}Q_{ai}Q_{bj} -  \dfrac{1}{4} \tilde{\mathcal{G}}^{ij}  \tilde{\mathcal{G}}_{ij} \bigg)  \notag \\
			&& + 2 \sigma  |\det(\tilde{\Pi}^{ai})|^{1/(n-2)} \Lambda.
		\end{eqnarray}
	\end{subequations}
Here $\Gamma_{a}{}^{i}{}_{j}$ is the connection compatible with $\tilde{\Pi}^{ai}$,
	\begin{equation}
		\nabla_{a}\tilde{\Pi}^{bi} :=  \partial_{a}\tilde{\Pi}^{bi} + \Gamma^{b}{}_{a c}\tilde{\Pi}^{ci} - \Gamma^{c}{}_{a c}\tilde{\Pi}^{bi} + \Gamma_{a}{}^{i}{}_{j}\tilde{\Pi}^{bj} = 0,
	\end{equation}
	and $R_{ab}{}^{i}{}_{j} = \partial_{a}\Gamma_{b}{}^{i}{}_{j} - \partial_{b} \Gamma_{a}{}^{i}{}_{j} + \Gamma_{a}{}^{i}{}_{k} \Gamma_{b}{}^{k}{}_{j} - \Gamma_{b}{}^{i}{}_{k} \Gamma_{a}{}^{k}{}_{j}$  is the curvature of $\Gamma_{a}{}^{i}{}_{j}$.

	Adding the Gauss constraint~\eqref{Gauss_def2} to the action~\eqref{S_ADM2} and factoring out all the terms proportional to this constraint, the action acquires the form
	\begin{eqnarray}
		\label{S_SO}
		S &= & \int_{\mathbb{R} \times \Sigma} dt d^{n-1}x  \Big( 2\tilde{\Pi}^{ai}\dot{Q}_{ai} -\lambda_{ij} \tilde{\mathcal{G}}^{ij} \notag \\
		&& - 2 N^{a}\tilde{\mathcal{D}}_{a} - \underaccent{\tilde}{N} \tilde{\tilde{\mathcal{H}}} \Big),
	\end{eqnarray}
	with $\uac{N} = |\det(\tilde{\Pi}^{ai})|^{-1/(n-2)} N$ and the first-class constraints are given by
	\begin{subequations}
		\begin{eqnarray}
			\label{G_SO}
			\tilde{\mathcal{G}}^{ij} &=& 2 \tilde{\Pi}^{a[i} Q_{a}{}^{j]}, \\
			\tilde{\mathcal{D}}_{a} &:=& 2 \tilde{\Pi}^{bi} \partial_{[a} Q_{b]i} - Q_{ai} \partial_{b}\tilde{\Pi}^{bi}, \\
			\label{H_SO}
			\tilde{\tilde{\mathcal{H}}} &:=& - \sigma \tilde{\Pi}^{ai} \tilde{\Pi}^{bj} R_{abij} +  2 \tilde{\Pi}^{a[i}\tilde{\Pi}^{|b|j]} Q_{ai} Q_{bj} \notag \\
			&& + 2 \sigma |\det(\tilde{\Pi}^{ai})|^{2/(n-2)} \Lambda,
		\end{eqnarray}
	\end{subequations}
and $\lambda_{ij} = -\lambda_{ji}$. Note that we neglected the boundary term in~\eqref{ss_ADM2} and also the boundary term that comes out from the integration by parts of $\nabla_{a} \tilde{\mathcal{G}}^{ij}$. 
	
	The formulation encompassed by the action~\eqref{S_SO} with the constraints~\eqref{G_SO}--\eqref{H_SO} is precisely the Hamiltonian formulation corresponding to the $SO(n-1)$ formulation derived from the $n$-dimensional Palatini action~\eqref{S_QP_tg}~\cite{Montesinos2001}.

\section{From ADM to Ashtekar-Barbero variables and vice versa}\label{Ashtekar_Barbero}

We can directly obtain the Barbero formulation~\cite{Barbero9505} of general relativity from the ADM formulation~\cite{ADM}. This new approach avoids the canonical transformation from the $SO(3)$ phase-space variables as was done by Barbero himself and is also an alternative to other Hamiltonian methods that get such a formulation from the Holst action~\cite{Holst9605,Barros0100,Montesinos1801,Montesinos1903,Montesinos2004a}. As the reader can guess, this (bottom-up) approach is analogous to the one presented in Sec.~\ref{ADM_to_SO}, and thus the starting point is precisely the ADM formulation~\eqref{getting_ADM} in a spacetime of dimension four, which is given by the action
\begin{equation}\label{ADM_4d}
        S =  \int_{\mathbb{R} \times \Sigma}dt d^{3}x\Big (  \tilde{\pi}^{ab} \dot{q}_{ab}  - 2 N^{a}\tilde{\mathcal{C}}_{a} - N \tilde{\mathcal{C}} \Big).
    \end{equation}

Similarly, it is possible to perform the top-down approach, i.e., to start from the Barbero formulation and to get the ADM one following the same ideas developed in Sec.~\ref{SO_to_ADM}. All of this is done in what follows.
	
	\subsection{From ADM to Ashtekar-Barbero variables}\label{Barbero}
We begin by introducing the densitized triad $\tilde{\Pi}^{ai}$ that is related to the inverse of the spatial metric $q_{ab}$ through 
	\begin{equation}
	    \label{q_AB_inv}
	    q^{ab} = \dfrac{\delta_{ij} \tilde{\Pi}^{ai} \tilde{\Pi}^{bj}}{\mid \det(\tilde{\Pi}^{ai})\mid} ,
	\end{equation}
	where $\det (\tilde{\Pi}^{ai})$ is a tensor density of weight $2$ and the $SO(3)$ indices $i,j,k,\ldots$ take on the values $1, 2, 3$ and are lowered (raised) with the three-dimensional Euclidean metric $\delta_{ij}$ ($\delta^{ij}$). Let $\uac{\Pi}_{ai}$ be the inverse of $\tilde{\Pi}^{ai}$, i.e., $\uac{\Pi}_{ai} \tilde{\Pi}^{aj}=\delta_{i}^{j}$ and $\uac{\Pi}_{ai} \tilde{\Pi}^{bi}=\delta_{a}^{b}$, so the spatial metric is given by
	\begin{equation}
	    \label{q_AB}
	    q_{ab} = \mid \det ( \tilde{\Pi}^{ai} ) \mid  \delta^{ij}\uac{\Pi}_{ai} \uac{\Pi}_{bj}.
	\end{equation}
	
	Next, we introduce the covariant derivative $\nabla_{a}$ compatible with the densitized triad $\tilde{\Pi}^{ai}$, i.e., 
	\begin{equation}
		\label{CD_AB}
		\nabla_{a}\tilde{\Pi}^{bi} :=  \partial_{a}\tilde{\Pi}^{bi} + \Gamma^{b}{}_{a c}\tilde{\Pi}^{ci} - \Gamma^{c}{}_{a c}\tilde{\Pi}^{bi} + \epsilon^{i}{}_{jk} \Gamma_{a}{}^{j} \tilde{\Pi}^{bk} = 0,
	\end{equation}
	where $\Gamma^{a}{}_{bc} = \Gamma^{a}{}_{cb}$. The previous relations allow us to get both $\Gamma^{a}{}_{bc}$ and $\Gamma_{ai}$ in terms of $\tilde{\Pi}^{ai}$ and their derivatives. Note that $\Gamma^{a}{}_{bc}$ is the connection constructed with  $q_{ab}$ and $\Gamma_{ai}$ is related to the $SO(3)$ connection $\Gamma_{a}{}^{i}{}_{j}$ used in Sec.~\ref{vectors_ADM} by $\Gamma_{ai}:= - (1/2) \epsilon_{ijk} \Gamma_{a}{}^{jk}$, where $\epsilon_{ijk}$ is the totally antisymmetric $SO(3)$ invariant tensor, with $\epsilon_{123}=1$.
	
	Now, we introduce the Ashtekar-Barbero connection $A_{ai}$ into the formalism through the relation
	\begin{eqnarray}
		\label{pi_AB}
		\tilde{\pi}^{ab} &=& \dfrac{|\det(\tilde{\Pi}^{ai})|}{\gamma}  \left(Q^{-1}\right)^{abcd} \uac{\Pi}_{(c}{}^{i} \left( A_{d)i} - \Gamma_{d)i} \right), 
	\end{eqnarray}
	where $\gamma$ is the Barbero-Immirzi parameter and
	\begin{eqnarray}
        \label{Q_def2}
        \left(Q^{-1}\right)^{abcd} &:=& \frac{1}{2} \left( q^{ac} q^{bd} + q^{ad} q^{bc}  - 2 q^{ab} q^{cd} \right), 
    \end{eqnarray}
	with $q^{ab}$ given by \eqref{q_AB_inv}.\footnote{Notice that $\left(Q^{-1}\right)^{abcd} =  |\det(\tilde{\Pi}^{ai})|^{-1/2} G^{abcd}$, where $G^{abcd}$ is the inverse of the so-called supermetric $G_{abcd}$~\cite{DeWitt6708}.} Therefore, the ADM variables~\eqref{q_AB} and~\eqref{pi_AB} have been parametrized in terms of the Ashtekar-Barbero variables $(A_{ai}, \tilde{\Pi}^{ai})$.
	
	To go from the phase space described by the ADM variables $(q_{ab}, \tilde{\pi}^{ab})$ to the Ashtekar-Barbero variables $(A_{ai}, \tilde{\Pi}^{ai})$, we need to introduce the Gauss constraint
	\begin{equation}
		\label{G_AB}
		\tilde{\mathcal{G}}^{i} := \dfrac{1}{\gamma} \left( \partial_{a} \tilde{\Pi}^{ai} + \epsilon^{i}{}_{jk} A_{a}{}^{j} \tilde{\Pi}^{ak} \right) \approx 0.
	\end{equation}
	
	Continuing with the analysis, we substitute \eqref{q_AB} and \eqref{pi_AB} in the symplectic structure of~\eqref{ADM_4d}, and we get
	\begin{eqnarray}
		\label{ss_AB}
		\tilde{\pi}^{ab} \dot{q}_{ab}  &=& \dfrac{2}{\gamma} \tilde{\Pi}^{ai} \dot{A}_{ai} +  \epsilon_{ijk}  \uac{\Pi}_{a}{}^{j} \tilde{\mathcal{G}}^{k} \partial_{t}\tilde{\Pi}^{ai} \nonumber\\
		&& - \dfrac{2}{\gamma} \partial_{t}\big[( A_{ai} - \Gamma_{ai}) \tilde{\Pi}^{ai}\big] \nonumber\\
		&& +\dfrac{1}{\gamma} \partial_{a} \left( \epsilon_{ijk} \tilde{\Pi}^{ai} \uac{\Pi}_{b}{}^{j} \partial_{t} \tilde{\Pi}^{bk} \right). 
	\end{eqnarray}
	Thus, modulo the Gauss constraint~\eqref{G_AB}, the variable $A_{ai}$ is canonically conjugated to $\tilde{\Pi}^{ai}$. On the other hand, using~\eqref{q_AB} and \eqref{pi_AB}, the diffeomorphism and Hamiltonian constraints that appear in~\eqref{ADM_4d} become
	\begin{subequations}
		\begin{eqnarray}
			\tilde{\mathcal{C}}_{a} &=& \dfrac{1}{\gamma}\left( 2\tilde{\Pi}^{bi} \partial_{[a} A_{b]i} - A_{ai} \partial_{b}\tilde{\Pi}^{bi} \right) +  \Gamma_{ai} \tilde{\mathcal{G}}^{i} \notag \\
			&& - \dfrac{1}{2} \epsilon_{ijk} \uac{\Pi}_{a}{}^{i} \tilde{\Pi}^{bj} \nabla_{b} \tilde{\mathcal{G}}^{k}, \label{Abhay_Ca}	\\
			\tilde{\mathcal{C}} &=& \dfrac{1}{\gamma^{2} |\det(\tilde{\Pi}^{ai})|^{1/2}} \epsilon_{ijk} \tilde{\Pi}^{ai} \tilde{\Pi}^{bj} \big[ F_{ab}{}^{k} \notag \\
			&&+ (\sigma \gamma^{2} -1) R_{ab}{}^{k} \big]  + 2 \sigma  |\det(\tilde{\Pi}^{ai})|^{1/2} \Lambda \notag \\
			&&+ \dfrac{1}{|\det(\tilde{\Pi}^{ai})|^{1/2}} \bigg( \dfrac{2}{\gamma } \tilde{\Pi}^{ai} \nabla_{a} \tilde{\mathcal{G}}_{i} - \dfrac{1}{2} \tilde{\mathcal{G}}^{i}  \tilde{\mathcal{G}}_{i} \bigg), \quad \label{Abhay_C}
		\end{eqnarray}
	\end{subequations}
	where we used $\epsilon_{ijk} \tilde{\Pi}^{ai} \tilde{\Pi}^{bj} R_{ab}{}^{k} = -qR$ and the identity
	\begin{eqnarray}
		\label{id}
		2 \tilde{\Pi}^{ai} \nabla_{a} \tilde{\mathcal{G}}_{i} &=& - \dfrac{1}{\gamma} \epsilon_{ijk} \tilde{\Pi}^{ai} \tilde{\Pi}^{bj} \Big[ F_{ab}{}^{k} - R_{ab}{}^{k} \notag \\
		&& - \epsilon^k{}_{lm} \left( A_a{}^l - \Gamma_a{}^l \right) \left( A_b{}^m - \Gamma_b{}^m \right) \Big], \quad
	\end{eqnarray}
	that relates the curvatures 
	\begin{subequations}
		\begin{eqnarray}
			F_{ab}{}^i &:=& \partial_{a} A_b{}^i - \partial_{b} A_a{}^i + \epsilon^i{}_{jk} A_{a}{}^{j} A_{b}{}^{k}, \\
			R_{ab}{}^i &:=& \partial_{a} \Gamma_b{}^i - \partial_{b} \Gamma_a{}^i + \epsilon^i{}_{jk} \Gamma_{a}{}^{j} \Gamma_{b}{}^{k},
		\end{eqnarray}
	\end{subequations}
	of $A_a{}^i$ and $\Gamma_a{}^i$, respectively.
		
	Therefore, by substituting ~\eqref{ss_AB},~\eqref{Abhay_Ca},~\eqref{Abhay_C} into the action~\eqref{ADM_4d}, adding the Gauss constraint~\eqref{G_AB} and after factoring out all the terms proportional to the Gauss constraint, the action~\eqref{ADM_4d} acquires the form
	\begin{eqnarray}
		\label{S_AB}
		S &= & \int_{\mathbb{R} \times \Sigma} dt d^{3}x  \Big( \dfrac{2}{\gamma} \tilde{\Pi}^{ai}\dot{A}_{ai} -\lambda_{i} \tilde{\mathcal{G}}^{i} - 2 N^{a}\tilde{\mathcal{D}}_{a} - \underaccent{\tilde}{N} \tilde{\tilde{\mathcal{H}}} \Big), \notag \\
	\end{eqnarray}
	where we redefined $\uac{N}:= |\det(\tilde{\Pi}^{ai})|^{-1/2} N$ and the first-class constraints are given by
	\begin{subequations}
		\begin{eqnarray}
			\label{G_AB2}
			\tilde{\mathcal{G}}^{i} &=&  \dfrac{1}{\gamma} \left( \partial_{a} \tilde{\Pi}^{ai} + \epsilon^{i}{}_{jk} A_{a}{}^{j} \tilde{\Pi}^{ak} \right), \\
			\tilde{\mathcal{D}}_{a} &:=& \dfrac{1}{\gamma} \left( 2 \tilde{\Pi}^{bi} \partial_{[a} A_{b]i} - A_{ai} \partial_{b}\tilde{\Pi}^{bi} \right), \\
			\label{H_AB}
			\tilde{\tilde{\mathcal{H}}} &:=& \dfrac{1}{\gamma^{2}} \epsilon_{ijk} \tilde{\Pi}^{ai} \tilde{\Pi}^{bj} \left[ F_{ab}{}^{k} + (\sigma \gamma^{2} -1) R_{ab}{}^{k} \right]  \notag \\
			&& + 2 \sigma  |\det(\tilde{\Pi}^{ai})| \Lambda.
		\end{eqnarray}
	\end{subequations}
	This is precisely the Barbero formulation~\cite{Barbero9505} of general relativity in four dimensions, which is the starting point in the loop quantum gravity approach~\cite{RovBook,ThieBook}. As far as we understand, the current approach is the easiest way of getting the Hamiltonian formulation of general relativity in terms of the Ashtekar-Barbero variables $A_{ai}$ and $\tilde{\Pi}^{ai}$. Note that our result~\eqref{S_AB} matches the conventions of Ref.~\cite{ThieBook}.

	\subsection{From Ashtekar-Barbero to ADM variables}
	For the sake of completeness, we now analyze the top-down approach, in which we go from the Barbero to the ADM formulation. Therefore, the starting point is the formulation given by the action~\cite{Barbero9505} (see also \cite{Holst9605,Barros0100,Montesinos1801,Montesinos1903,Montesinos2004a})
		\begin{eqnarray}
		\label{S_AB_2}
		S &= & \int_{\mathbb{R} \times \Sigma} \!\!\!\!\!\! dt d^{3}x  \left ( \dfrac{2}{\gamma} \tilde{\Pi}^{ai}\dot{A}_{ai} -\lambda_{i} \tilde{\mathcal{G}}^{i} - 2 N^{a}\tilde{\mathcal{D}}_{a} - \underaccent{\tilde}{N} \tilde{\tilde{\mathcal{H}}} \right ), \notag  \\ 
	\end{eqnarray}
	where the first-class constraints are given by
	\begin{subequations}
		\begin{eqnarray}
			\label{G_AB_2}
			\tilde{\mathcal{G}}^{i} &=&  \dfrac{1}{\gamma} \left( \partial_{a} \tilde{\Pi}^{ai} + \epsilon^{i}{}_{jk} A_{a}{}^{j} \tilde{\Pi}^{ak} \right), \\
			\tilde{\mathcal{D}}_{a} &:=& \dfrac{1}{\gamma} \left( 2 \tilde{\Pi}^{bi} \partial_{[a} A_{b]i} - A_{ai} \partial_{b}\tilde{\Pi}^{bi} \right), \label{Abhay_difeos} \\
			\label{H_AB_2}
			\tilde{\tilde{\mathcal{H}}} &:=& \dfrac{1}{\gamma^{2}} \epsilon_{ijk} \tilde{\Pi}^{ai} \tilde{\Pi}^{bj} \left[ F_{ab}{}^{k} + (\sigma \gamma^{2} -1) R_{ab}{}^{k} \right]  \notag \\
			&& + 2 \sigma  |\det(\tilde{\Pi}^{ai})| \Lambda.
		\end{eqnarray}
	\end{subequations}
	Here, $F_{abi}$ and $R_{abi}$ are the curvatures of the corresponding connections $A_{ai}$ and $\Gamma_{ai}$.
	
	In the Barbero formulation, the phase-space variables are the $SO(3)$ connection $A_{ai}$ and the densitized triad $\tilde{\Pi}^{ai}$. To get the ADM formulation from the Barbero one, we need to define the ADM variables in terms of the Ashtekar-Barbero variables
	\begin{subequations}
		\begin{eqnarray}
			\label{q_def_AB}
			q_{ab} &:=& | \det(\tilde{\Pi}^{ai}) | \uac{\Pi}_{ai} \uac{\Pi}_{b}{}^{i}, \\
			\label{pi_def_AB}
			\tilde{\pi}^{ab} &:=& \dfrac{1}{\gamma} |\det(\tilde{\Pi}^{ai})| \left( Q^{-1} \right)^{abcd} \uac{\Pi}_{(c}{}^{i} \left( A_{d)i} - \Gamma_{d)i} \right), \quad \quad
		\end{eqnarray}
	\end{subequations}
	where $\uac{\Pi}_{ai}$ is the inverse of $\tilde{\Pi}^{ai}$, so $q^{ab} = | \det(\tilde{\Pi}^{ai}) |^{-1} \tilde{\Pi}^{ai} \tilde{\Pi}^{b}{}_{i}$ is the inverse of $q_{ab}$. Here, $\left( Q^{-1} \right)^{abcd}$ is given by~\eqref{Q_def2}.
	
Note that the expression~\eqref{pi_def_AB} can be thought as six linear equations for nine unknowns $A_{ai}$. With this in mind, we can solve for $A_{ai}$ and obtain 
	\begin{eqnarray}
		\label{A2pi}
		A_{ai} &=& \Gamma_{ai} + \gamma |\det(\tilde{\Pi}^{ai})|^{-1} Q_{abcd} \tilde{\Pi}^{b}{}_{i} \tilde{\pi}^{cd} \notag \\
		&& + \dfrac{\gamma}{2} \epsilon_{ijk} \uac{\Pi}_{a}{}^{j} \tilde{\Pi}^{k},
	\end{eqnarray}
with 
\begin{eqnarray}
\label{Q_inv_AB}
 Q_{abcd} &:=& \dfrac{1}{2} \left (q_{ac} q_{bd} + q_{ad} q_{bc} - q_{ab} q_{cd} \right),
\end{eqnarray}
where $Q_{abef} \left( Q^{-1}\right)^{efcd} = (1/2) \left( \delta^{c}_{a} \delta^{d}_{b} + \delta^{d}_{a} \delta^{c}_{b} \right)$. Thus, the variables  $\tilde{\Pi}^{i}$ that appear in the right-hand side of~\eqref{A2pi} are three independent variables to account for the difference in the number of independent variables between $A_{ai}$ and $\tilde{\pi}^{ab}$.
	
	Therefore, using~\eqref{A2pi} the symplectic structure of~\eqref{S_AB_2} can be written as
	\begin{eqnarray}\label{ss_AB_ADM}
			\dfrac{2}{\gamma} \tilde{\Pi}^{ai} \dot{A}_{ai}  &=& \tilde{\pi}^{ab} \dot{q}_{ab} - \epsilon_{ijk}  \uac{\Pi}_{a}{}^{j} \tilde{\Pi}^{k} \partial_{t}\tilde{\Pi}^{ai}  \nonumber\\
			&& + \dfrac{2}{\gamma} \partial_{t}\big[( A_{ai} - \Gamma_{ai}) \tilde{\Pi}^{ai}\big] \nonumber\\
			&& -\dfrac{1}{\gamma} \partial_{a} \left( \epsilon_{ijk} \tilde{\Pi}^{ai} \uac{\Pi}_{b}{}^{j} \partial_{t} \tilde{\Pi}^{bk} \right).
	\end{eqnarray}
	On the other hand, the first-class constraints~\eqref{G_AB_2},~\eqref{Abhay_difeos}, and~\eqref{H_AB_2} become
	\begin{subequations}
		\begin{eqnarray}
			\label{G_AB_ADM}
			\tilde{\mathcal{G}}^{i} &=& \tilde{\Pi}^{i}, \\
			\label{Abhay_difeos_2}
			\tilde{\mathcal{D}}_{a} &=& - q_{ab} D_{c} \tilde{\pi}^{bc}  - \Gamma_{ai} \tilde{\Pi}^{i} + \dfrac{1}{2} \epsilon_{ijk} \uac{\Pi}_{a}{}^{i} \tilde{\Pi}^{bj} \nabla_{b} \tilde{\Pi}^{k}, \quad \quad \\
			\label{H_missing}
			\tilde{\tilde{\mathcal{H}}} &=& - \sigma qR  - Q_{abcd}\tilde{\pi}^{ab} \tilde{\pi}^{cd} + 2 \sigma q \Lambda \notag \\
			&& - \dfrac{2}{\gamma} \tilde{\Pi}^{ai} \nabla_{a} \tilde{\Pi}_{i}+ \dfrac{1}{2} \tilde{\Pi}^{i} \tilde{\Pi}_{i},
		\end{eqnarray}
	\end{subequations}
	where $q=\det{(q_{ab})}$, $R := q^{ac} q^{bd} R_{abcd}$ is the scalar curvature of the Levi-Civita connection $\Gamma^{a}{}_{bc}$, and $D_{a}$ is the covariant derivative with respect to such a connection [see Eq.~\eqref{Dp}].  We also used the identities~\eqref{id} and $\epsilon_{ijk} \tilde{\Pi}^{ai} \tilde{\Pi}^{bj} R_{ab}{}^{k} = -qR$.
	
	By substituting~\eqref{ss_AB_ADM},~\eqref{G_AB_ADM},~\eqref{Abhay_difeos_2}, and~\eqref{H_missing} into the action~\eqref{S_AB_2}, and after factoring out all the terms proportional to the Gauss constraint, which implies replacing the Lagrange multiplier $\lambda_i$ with $\Lambda_i$, and neglecting the boundary terms in~\eqref{ss_AB_ADM}, the action~\eqref{S_AB_2} acquires the form 
	\begin{eqnarray}
		S &= & \int_{\mathbb{R} \times \Sigma}dt d^{3}x\Big (  \tilde{\pi}^{ab} \dot{q}_{ab} - 2 N^{a}\tilde{\mathcal{C}}_{a} - N \tilde{\mathcal{C}} -\Lambda_{i} \tilde{\mathcal{G}}^{i} \Big ), \notag \\
	\end{eqnarray}
	where $N= \sqrt{q} \uac{N}$, $\tilde{\mathcal{G}}^{i}$ is given by~\eqref{G_AB_ADM} and the diffeomorphism and Hamiltonian constraints are 
	\begin{subequations}
		\begin{eqnarray}
			\label{Ca_AB}
			\tilde{\mathcal{C}}_{a} &:=& - q_{ab} D_{c} \tilde{\pi}^{bc}, \\
			\label{C_AB}
			\tilde{\mathcal{C}} &:=& - \sigma \sqrt{q} R  - \dfrac{1}{\sqrt{q}} Q_{abcd}\tilde{\pi}^{ab} \tilde{\pi}^{cd} + 2 \sigma \sqrt{q} \Lambda.
		\end{eqnarray}
	\end{subequations}
	Therefore, just as in Secs.~\ref{Lorentz_to_ADM} and \ref{SO_to_ADM}, the ADM formulation lies in the hypersurface where the Gauss constraint is satisfied $\tilde{\mathcal{G}}^{i}=0$. To provide a rigorous Hamiltonian formulation~\cite{DiracBook}, we must introduce the variable $Y_{i}$ as the configuration variable conjugated to $\tilde{\Pi}^{i}$. Then, we add the term $\tilde{\Pi}^{i} \dot{Y}_{i}$ and the corresponding constraint $Y_{i} \approx 0$. Thus, the theory is described by the action
	\begin{eqnarray}
		S &= & \int_{\mathbb{R} \times \Sigma}dt d^{3}x\Big ( \tilde{\pi}^{ab} \dot{q}_{ab} + \tilde{\Pi}^{i} \dot{Y}_{i} - 2 N^{a}\tilde{\mathcal{C}}_{a} - N \tilde{\mathcal{C}}  \notag \\
		&& -\Lambda_{i} \tilde{\mathcal{G}}^{i}  - \tilde{\rho}^{i} Y_{i} \Big), 
	\end{eqnarray}
	where $\tilde{\rho}^{i}$ is the Lagrange multiplier that imposes the constraint $Y_{i} \approx 0$. Since $\tilde{\mathcal{G}}^{i} \approx 0$ and $Y_{i} \approx 0$ are second-class constraints (with a possible modification of $\tilde{\mathcal{C}}_{a}$ and $\tilde{\mathcal{C}} $ involving the second-class constraints), we make them strongly equal to zero and obtain the ADM formulation of general relativity given by the Hamiltonian action principle~\cite{ADM}
	\begin{equation}
		S =  \int_{\mathbb{R} \times \Sigma}dt d^{3}x\Big (  \tilde{\pi}^{ab} \dot{q}_{ab}  - 2 N^{a}\tilde{\mathcal{C}}_{a} - N \tilde{\mathcal{C}} \Big),
	\end{equation}
	where $(q_{ab}, \tilde{\pi}^{ab})$ are the canonical variables and the constraints are given by \eqref{Ca_AB} and \eqref{C_AB}.

	\section{Concluding remarks}\label{Sec_concl}
	We conclude the paper by making some remarks. Many researchers are familiar with the ADM formulation of general relativity~\cite{ADM} (see also~\cite{ThieBook})---for instance, those working in numerical relativity---but not with the Hamiltonian formulations of general relativity that come from Lagrangian actions that depend functionally on the vielbein and the connection, such as the Palatini or Holst actions~\cite{Holst9605}. Such readers can now use the approach of the Sec.~\ref{Ashtekar_Barbero} of this paper to go immediately to the Hamiltonian formulation of general relativity in terms of Ashtekar--Barbero variables~\cite{Barbero9505} as an alternative approach to the several ways of getting such a formulation~\cite{Barbero9505,Holst9605,Barros0100,Montesinos1801,Montesinos1903,Montesinos2004a}. Similarly, using the approach of Secs.~\ref{ADM_to_Palatini} and~\ref{vectors_ADM}, readers can reach directly the $SO(n-1,1)$ [$SO(n)$] and $SO(n-1)$ Hamiltonian formulations of general relativity described by the Palatini action~\cite{Montesinos2001,Bodendorfer1301a,Bodendorfer1301b} starting only from the ADM formulation. These facts are relevant and can be used, for instance, to express any previous result based on ADM variables---such as conserved quantities---in terms of the phase-space variables involved in the other Hamiltonian formulations. We think all these results fill out a gap present in the literature on the canonical analysis of general relativity. Furthermore, the reduction process reported in Secs.~\ref{Lorentz_to_ADM}, \ref{vectors_ADM} and~\ref{Ashtekar_Barbero} of this paper from any of the first-order Hamiltonian formulations of general relativity to the ADM formulation can also be viewed in this way, namely, as a fast track that circumvents the subtleties and technical details of the canonical analysis and allows us to reach the standard ADM formulation. \\

	\acknowledgments
We thank the valuable comments of Mariano Celada, Ricardo Escobedo, Alejandro Perez,  and José David Vergara. This work was partially supported by Consejo Nacional de Ciencia y Tecnología (CONACyT), M\'{e}xico, Grant No. A1-S-7701.


	\bibliographystyle{apsrev4-1}
	
	\bibliography{References}

\begin{thebibliography}{15}%
\makeatletter
\providecommand \@ifxundefined [1]{%
 \@ifx{#1\undefined}
}%
\providecommand \@ifnum [1]{%
 \ifnum #1\expandafter \@firstoftwo
 \else \expandafter \@secondoftwo
 \fi
}%
\providecommand \@ifx [1]{%
 \ifx #1\expandafter \@firstoftwo
 \else \expandafter \@secondoftwo
 \fi
}%
\providecommand \natexlab [1]{#1}%
\providecommand \enquote  [1]{``#1''}%
\providecommand \bibnamefont  [1]{#1}%
\providecommand \bibfnamefont [1]{#1}%
\providecommand \citenamefont [1]{#1}%
\providecommand \href@noop [0]{\@secondoftwo}%
\providecommand \href [0]{\begingroup \@sanitize@url \@href}%
\providecommand \@href[1]{\@@startlink{#1}\@@href}%
\providecommand \@@href[1]{\endgroup#1\@@endlink}%
\providecommand \@sanitize@url [0]{\catcode `\\12\catcode `\$12\catcode
  `\&12\catcode `\#12\catcode `\^12\catcode `\_12\catcode `\%12\relax}%
\providecommand \@@startlink[1]{}%
\providecommand \@@endlink[0]{}%
\providecommand \url  [0]{\begingroup\@sanitize@url \@url }%
\providecommand \@url [1]{\endgroup\@href {#1}{\urlprefix }}%
\providecommand \urlprefix  [0]{URL }%
\providecommand \Eprint [0]{\href }%
\providecommand \doibase [0]{http://dx.doi.org/}%
\providecommand \selectlanguage [0]{\@gobble}%
\providecommand \bibinfo  [0]{\@secondoftwo}%
\providecommand \bibfield  [0]{\@secondoftwo}%
\providecommand \translation [1]{[#1]}%
\providecommand \BibitemOpen [0]{}%
\providecommand \bibitemStop [0]{}%
\providecommand \bibitemNoStop [0]{.\EOS\space}%
\providecommand \EOS [0]{\spacefactor3000\relax}%
\providecommand \BibitemShut  [1]{\csname bibitem#1\endcsname}%
\let\auto@bib@innerbib\@empty
\bibitem [{\citenamefont {Arnowitt}\ \emph {et~al.}(2008)\citenamefont
  {Arnowitt}, \citenamefont {Deser},\ and\ \citenamefont {Misner}}]{ADM}%
  \BibitemOpen
  \bibfield  {author} {\bibinfo {author} {\bibfnamefont {R.}~\bibnamefont
  {Arnowitt}}, \bibinfo {author} {\bibfnamefont {S.}~\bibnamefont {Deser}}, \
  and\ \bibinfo {author} {\bibfnamefont {C.~W.}\ \bibnamefont {Misner}},\
  }\href {\doibase 10.1007/s10714-008-0661-1} {\bibfield  {journal} {\bibinfo
  {journal} {Gen. Relativ. Gravit.}\ }\textbf {\bibinfo {volume} {40}},\
  \bibinfo {pages} {1997} (\bibinfo {year} {2008})}\BibitemShut {NoStop}%
\bibitem [{\citenamefont {Dirac}(1964)}]{DiracBook}%
  \BibitemOpen
  \bibfield  {author} {\bibinfo {author} {\bibfnamefont {P.~A.~M.}\
  \bibnamefont {Dirac}},\ }\href@noop {} {\emph {\bibinfo {title} {Lectures on
  Quantum Mechanics}}}\ (\bibinfo  {publisher} {Belfer Graduate School of
  Science, Yeshiva University, New York},\ \bibinfo {year} {1964})\BibitemShut
  {NoStop}%
\bibitem [{\citenamefont {Barbero~G.}(1995)}]{Barbero9505}%
  \BibitemOpen
  \bibfield  {author} {\bibinfo {author} {\bibfnamefont {J.~F.}\ \bibnamefont
  {Barbero~G.}},\ }\href {\doibase 10.1103/PhysRevD.51.5507} {\bibfield
  {journal} {\bibinfo  {journal} {Phys. Rev. D}\ }\textbf {\bibinfo {volume}
  {51}},\ \bibinfo {pages} {5507} (\bibinfo {year} {1995})}\BibitemShut
  {NoStop}%
\bibitem [{\citenamefont {Rovelli}(2004)}]{RovBook}%
  \BibitemOpen
  \bibfield  {author} {\bibinfo {author} {\bibfnamefont {C.}~\bibnamefont
  {Rovelli}},\ }\href@noop {} {\emph {\bibinfo {title} {Quantum Gravity}}}\
  (\bibinfo  {publisher} {Cambridge University Press, Cambridge, England},\
  \bibinfo {year} {2004})\BibitemShut {NoStop}%
\bibitem [{\citenamefont {Thiemann}(2007)}]{ThieBook}%
  \BibitemOpen
  \bibfield  {author} {\bibinfo {author} {\bibfnamefont {T.}~\bibnamefont
  {Thiemann}},\ }\href@noop {} {\emph {\bibinfo {title} {Modern Canonical
  Quantum General Relativity}}}\ (\bibinfo  {publisher} {Cambridge University
  Press, Cambridge, England},\ \bibinfo {year} {2007})\BibitemShut {NoStop}%
\bibitem [{\citenamefont {Ashtekar}(1986)}]{Ashtekar8611}%
  \BibitemOpen
  \bibfield  {author} {\bibinfo {author} {\bibfnamefont {A.}~\bibnamefont
  {Ashtekar}},\ }\href {\doibase 10.1103/PhysRevLett.57.2244} {\bibfield
  {journal} {\bibinfo  {journal} {Phys. Rev. Lett.}\ }\textbf {\bibinfo
  {volume} {57}},\ \bibinfo {pages} {2244} (\bibinfo {year}
  {1986})}\BibitemShut {NoStop}%
\bibitem [{\citenamefont {Bodendorfer}\ \emph
  {et~al.}(2013{\natexlab{a}})\citenamefont {Bodendorfer}, \citenamefont
  {Thiemann},\ and\ \citenamefont {Thurn}}]{Bodendorfer1301b}%
  \BibitemOpen
  \bibfield  {author} {\bibinfo {author} {\bibfnamefont {N.}~\bibnamefont
  {Bodendorfer}}, \bibinfo {author} {\bibfnamefont {T.}~\bibnamefont
  {Thiemann}}, \ and\ \bibinfo {author} {\bibfnamefont {A.}~\bibnamefont
  {Thurn}},\ }\href {\doibase 10.1088/0264-9381/30/4/045002} {\bibfield
  {journal} {\bibinfo  {journal} {Classical Quantum Gravity}\ }\textbf
  {\bibinfo {volume} {30}},\ \bibinfo {pages} {045002} (\bibinfo {year}
  {2013}{\natexlab{a}})}\BibitemShut {NoStop}%
\bibitem [{\citenamefont {Montesinos}\ \emph
  {et~al.}(2020{\natexlab{a}})\citenamefont {Montesinos}, \citenamefont
  {Escobedo}, \citenamefont {Romero},\ and\ \citenamefont
  {Celada}}]{Montesinos2001}%
  \BibitemOpen
  \bibfield  {author} {\bibinfo {author} {\bibfnamefont {M.}~\bibnamefont
  {Montesinos}}, \bibinfo {author} {\bibfnamefont {R.}~\bibnamefont
  {Escobedo}}, \bibinfo {author} {\bibfnamefont {J.}~\bibnamefont {Romero}}, \
  and\ \bibinfo {author} {\bibfnamefont {M.}~\bibnamefont {Celada}},\ }\href
  {\doibase 10.1103/PhysRevD.101.024042} {\bibfield  {journal} {\bibinfo
  {journal} {Phys. Rev. D}\ }\textbf {\bibinfo {volume} {101}},\ \bibinfo
  {pages} {024042} (\bibinfo {year} {2020}{\natexlab{a}})}\BibitemShut
  {NoStop}%
\bibitem [{\citenamefont {Holst}(1996)}]{Holst9605}%
  \BibitemOpen
  \bibfield  {author} {\bibinfo {author} {\bibfnamefont {S.}~\bibnamefont
  {Holst}},\ }\href {\doibase 10.1103/PhysRevD.53.5966} {\bibfield  {journal}
  {\bibinfo  {journal} {Phys. Rev. D}\ }\textbf {\bibinfo {volume} {53}},\
  \bibinfo {pages} {5966} (\bibinfo {year} {1996})}\BibitemShut {NoStop}%
\bibitem [{\citenamefont {Barros~e S\'{a}}(2001)}]{Barros0100}%
  \BibitemOpen
  \bibfield  {author} {\bibinfo {author} {\bibfnamefont {N.}~\bibnamefont
  {Barros~e S\'{a}}},\ }\href {\doibase 10.1142/S0218271801000858} {\bibfield
  {journal} {\bibinfo  {journal} {Int. J. Mod. Phys. D}\ }\textbf {\bibinfo
  {volume} {10}},\ \bibinfo {pages} {261} (\bibinfo {year} {2001})}\BibitemShut
  {NoStop}%
\bibitem [{\citenamefont {Montesinos}\ \emph {et~al.}(2018)\citenamefont
  {Montesinos}, \citenamefont {Romero},\ and\ \citenamefont
  {Celada}}]{Montesinos1801}%
  \BibitemOpen
  \bibfield  {author} {\bibinfo {author} {\bibfnamefont {M.}~\bibnamefont
  {Montesinos}}, \bibinfo {author} {\bibfnamefont {J.}~\bibnamefont {Romero}},
  \ and\ \bibinfo {author} {\bibfnamefont {M.}~\bibnamefont {Celada}},\ }\href
  {\doibase 10.1103/PhysRevD.97.024014} {\bibfield  {journal} {\bibinfo
  {journal} {Phys. Rev. D}\ }\textbf {\bibinfo {volume} {97}},\ \bibinfo
  {pages} {024014} (\bibinfo {year} {2018})}\BibitemShut {NoStop}%
\bibitem [{\citenamefont {Montesinos}\ \emph {et~al.}(2019)\citenamefont
  {Montesinos}, \citenamefont {Romero},\ and\ \citenamefont
  {Celada}}]{Montesinos1903}%
  \BibitemOpen
  \bibfield  {author} {\bibinfo {author} {\bibfnamefont {M.}~\bibnamefont
  {Montesinos}}, \bibinfo {author} {\bibfnamefont {J.}~\bibnamefont {Romero}},
  \ and\ \bibinfo {author} {\bibfnamefont {M.}~\bibnamefont {Celada}},\ }\href
  {\doibase 10.1103/PhysRevD.99.064029} {\bibfield  {journal} {\bibinfo
  {journal} {Phys. Rev. D}\ }\textbf {\bibinfo {volume} {99}},\ \bibinfo
  {pages} {064029} (\bibinfo {year} {2019})}\BibitemShut {NoStop}%
\bibitem [{\citenamefont {Montesinos}\ \emph
  {et~al.}(2020{\natexlab{b}})\citenamefont {Montesinos}, \citenamefont
  {Romero},\ and\ \citenamefont {Celada}}]{Montesinos2004a}%
  \BibitemOpen
  \bibfield  {author} {\bibinfo {author} {\bibfnamefont {M.}~\bibnamefont
  {Montesinos}}, \bibinfo {author} {\bibfnamefont {J.}~\bibnamefont {Romero}},
  \ and\ \bibinfo {author} {\bibfnamefont {M.}~\bibnamefont {Celada}},\ }\href
  {\doibase 10.1103/PhysRevD.101.084003} {\bibfield  {journal} {\bibinfo
  {journal} {Phys. Rev. D}\ }\textbf {\bibinfo {volume} {101}},\ \bibinfo
  {pages} {084003} (\bibinfo {year} {2020}{\natexlab{b}})}\BibitemShut
  {NoStop}%
\bibitem [{\citenamefont {Bodendorfer}\ \emph
  {et~al.}(2013{\natexlab{b}})\citenamefont {Bodendorfer}, \citenamefont
  {Thiemann},\ and\ \citenamefont {Thurn}}]{Bodendorfer1301a}%
  \BibitemOpen
  \bibfield  {author} {\bibinfo {author} {\bibfnamefont {N.}~\bibnamefont
  {Bodendorfer}}, \bibinfo {author} {\bibfnamefont {T.}~\bibnamefont
  {Thiemann}}, \ and\ \bibinfo {author} {\bibfnamefont {A.}~\bibnamefont
  {Thurn}},\ }\href {\doibase 10.1088/0264-9381/30/4/045001} {\bibfield
  {journal} {\bibinfo  {journal} {Classical Quantum Gravity}\ }\textbf
  {\bibinfo {volume} {30}},\ \bibinfo {pages} {045001} (\bibinfo {year}
  {2013}{\natexlab{b}})}\BibitemShut {NoStop}%
\bibitem [{\citenamefont {DeWitt}(1967)}]{DeWitt6708}%
  \BibitemOpen
  \bibfield  {author} {\bibinfo {author} {\bibfnamefont {B.~S.}\ \bibnamefont
  {DeWitt}},\ }\href {\doibase 10.1103/PhysRev.160.1113} {\bibfield  {journal}
  {\bibinfo  {journal} {Phys. Rev.}\ }\textbf {\bibinfo {volume} {160}},\
  \bibinfo {pages} {1113} (\bibinfo {year} {1967})}\BibitemShut {NoStop}%
\end{thebibliography}%
	
\end{document}